# Dielectric losses in multi-layer Josephson junction qubits

David Gunnarsson[1], Juha-Matti Pirkkalainen[2], Jian Li[2], Gheorghe Sorin Paraoanu[2], Pertti Hakonen[2], Mika Sillanpää[2], and Mika Prunnila[1].

[1] VTT Technical Research Centre of Finland, P.O. Box 1000, FI-02044 VTT Espoo, Finland

[2] Low Temperature Laboratory, Aalto University School of Science, Espoo, Finland

E-mail: david.gunnarsson@vtt.fi

**Abstract**
We have measured the excited state lifetimes in Josephson junction phase and transmon qubits, all of which were fabricated with the same scalable multi-layer process. We have compared the lifetimes of phase qubits before and after removal of the isolating dielectric, $SiN_x$, and find a four-fold improvement of the relaxation time after the removal. Together with the results from the transmon qubit and measurements on coplanar waveguide resonators, these measurements indicate that the lifetimes are limited by losses from the dielectric constituents of the qubits. We have extracted the individual loss contributions from the dielectrics in the tunnel junction barrier, $AlO_x$, the isolating dielectric, $SiN_x$, and the substrate, $Si/SiO_2$, by weighing the total loss with the parts of electric field over the different dielectric materials. Our results agree well and complement the findings from other studies, demonstrating that superconducting qubits can be used as a reliable tool for high-frequency characterization of dielectric materials. We conclude with a discussion of how changes in design and material choice could improve qubit lifetimes up to a factor of four.

## 1. Introduction

A great deal of research has recently been dedicated to scaling up quantum computers from a few qubit regime to networks of quantum bits. These efforts have concentrated both on creating more robust qubit designs as well as better controlling the environmental non-idealities. Josephson junction (JJ) qubits, which include phase qubits [1, 2], flux qubits [3], transmon [4, 5] , quantronium [6], and charge qubits [7, 8] , are contenders in this effort of increasing the complexity of quantum electronics. In the first JJ qubit experiments, the coherence times were a few ns [6] and today, around one decade later, dozens of µs have been reported [9]. This impressive development comes from improvements in circuit design, with new ideas for devices [4, 6, 10, 11] and from thorough material research [12, 13, 14, 15].

In the perspective of reaching a scalable solution for multiple JJ qubits, fabrication by multi-layer technology seems preferable. For these qubits, the isolating material between the layers and the large tunnel junction sizes increases the coupling to unwanted degrees of freedom. A lot of work has been done to investigate these decoherence mechanisms and to quantify the quality of materials [12-16]. One limitation for these qubits originates from dielectric losses due to two-level systems (TLS), acting as electric dipoles [12, 16, 17]. In the frequency regime of multi-layer JJ qubits, 1-10 GHz, electric losses in the dielectrics are dominated by dipole relaxations. The dissipation can be quantified and measured through the loss factor, tan $\delta$, which can also be used to quantify the lifetimes of qubits that are limited by relaxation to TLS [12].

In this paper, we quantify the losses resulting from dielectric layers in two different qubit systems, a phase and a transmon qubit, made within the same multi-layer process. We have used their respective excited state lifetimes, combined with measurements of the photon lifetimes in coplanar waveguide (CPW) resonators, to extract the dielectric losses of the materials used at ultra-low temperature and with small probe powers. Compared to similar studies, we have also correlated the reduction of dielectric materials through sample post-processing to largely improve the qubit lifetime. It is worth noting, that often the quality of devices goes down when more steps are added to a process. Measurements from two different qubit types and CPWs have been used to quantify the quality of a highly scalable production process [18].

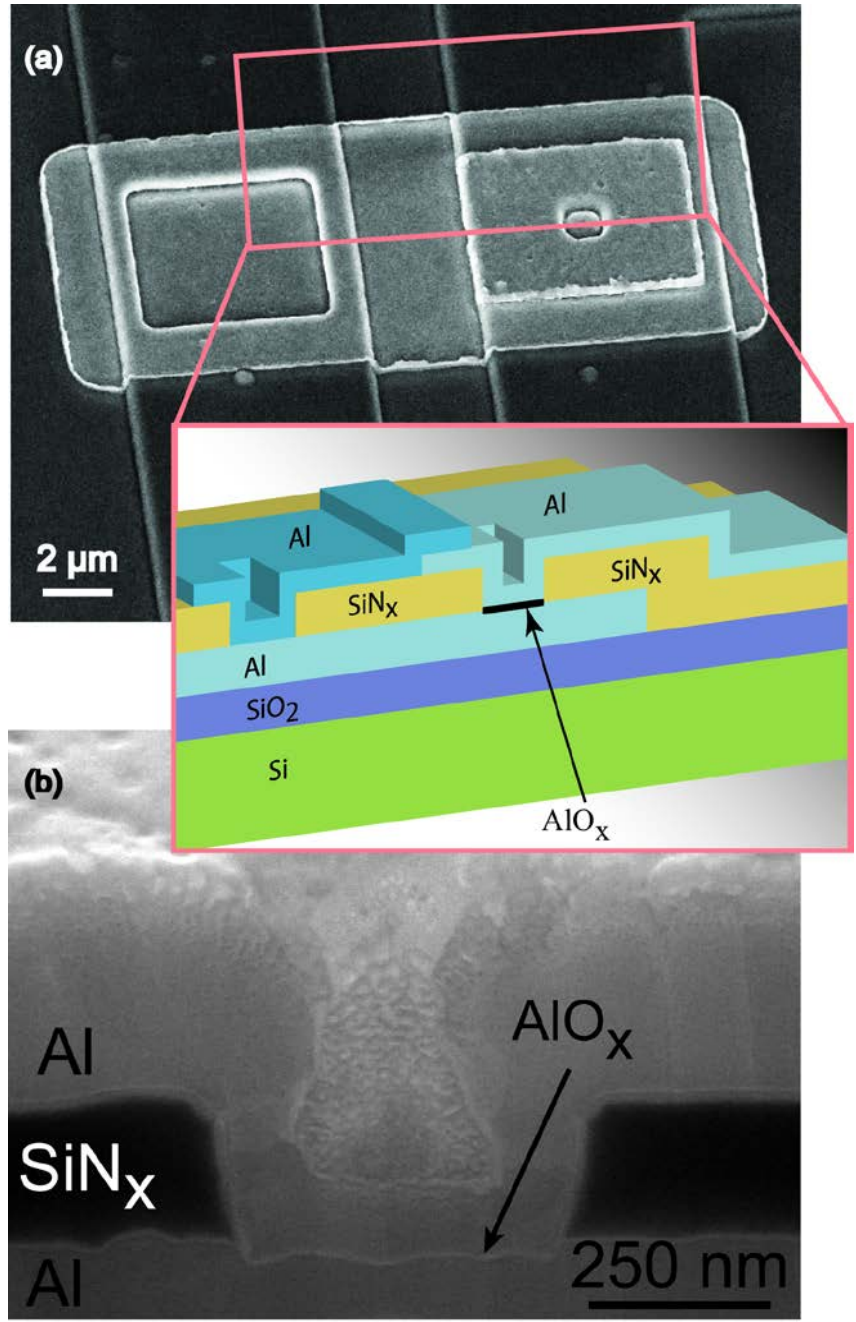

**Figure 1**. (a) SEM micrograph top view shows the tunnel junction (right) and the via (left). The inset illustrates the multi-layers and via in an illustrated cut view. (b) A SEM micrograph after Focused Ion Beam cut through the tunnel junction.

## 2. Experimental

Both Josephson junction phase qubits and transmon qubits [1-2, 4-5] were fabricated on thermally oxidized 625 µm-thick, 50 Ω-cm p-type Si substrates. The 300 nm-thick thermal oxide was grown at 1000 °C by wet oxidation. Tunnel junctions of the qubits have $AlO_x$ tunnel barriers and sputter deposited Al top and bottom electrodes (see figure 1 (a) and (b)). Tunnel junction area is defined by a through dielectric via process described and characterized in detail in Refs. [17]. We followed the process of Refs. [18] except that here we used 250 nm SiN$_x$ dielectric (deposited by plasma enhanced chemical vapour deposition at 180 °C). All patterns have been defined using UV-lithography. The transmon Josephson junctions were configured in a SQUID geometry with a nominal junction area $A_j$ =0.5 µm$^2$ (see figure 2 (a) and (b)), shunted with an Inter Digital Capacitor (IDC), $C_s^t$, and embedded into a CPW resonator designed for $f_c \simeq 5$ GHz (for device parameters, see table 1). The charging energy is defined through, $E_C = e^2/2C_\Sigma$ , where $C_\Sigma$ is the qubit total capacitance [4]. The phase qubits were made with the same technology, with the addition of a metallic bridge layer for galvanic contact between bottom and top Al layers. This adds two process steps, definition of vias in the SiN$_x$ and a bridge metallization of Al through lift-off (see figure 1 (a)). The phase qubits were made with the Josephson junction in a RF-SQUID configuration, with a nominal junction area of 1.7 µm$^2$, and coupled to an asymmetric DC-SQUID for read-out [2] (see figure 2 (c) and (d)). The Josephson junction was shunted with an IDC, $C_s^{PQ}$. Two different phase qubits were measured, PQ-A and PQ-B. For sample PQ-A the SiN$_x$ is left, while for sample PQ-B a final etch step with CF$_4$+O$_2$ plasma was applied for complete removal of the SiN$_x$ covering the IDC and surrounding the tunnel junction area. The removal gives rise to the change in total capacitance. The phase qubits are also coupled to $f_c = 6$ GHz CPW resonators, but they were far-detuned and not considered in these studies. Even though we compare two JJ qubits from different regimes (phase and intermediate charge-flux qubits), their tunnel junction barriers have similar physical dimensions, hence they can be regarded as similar in a material point of view and in the analysis of the dielectric losses.

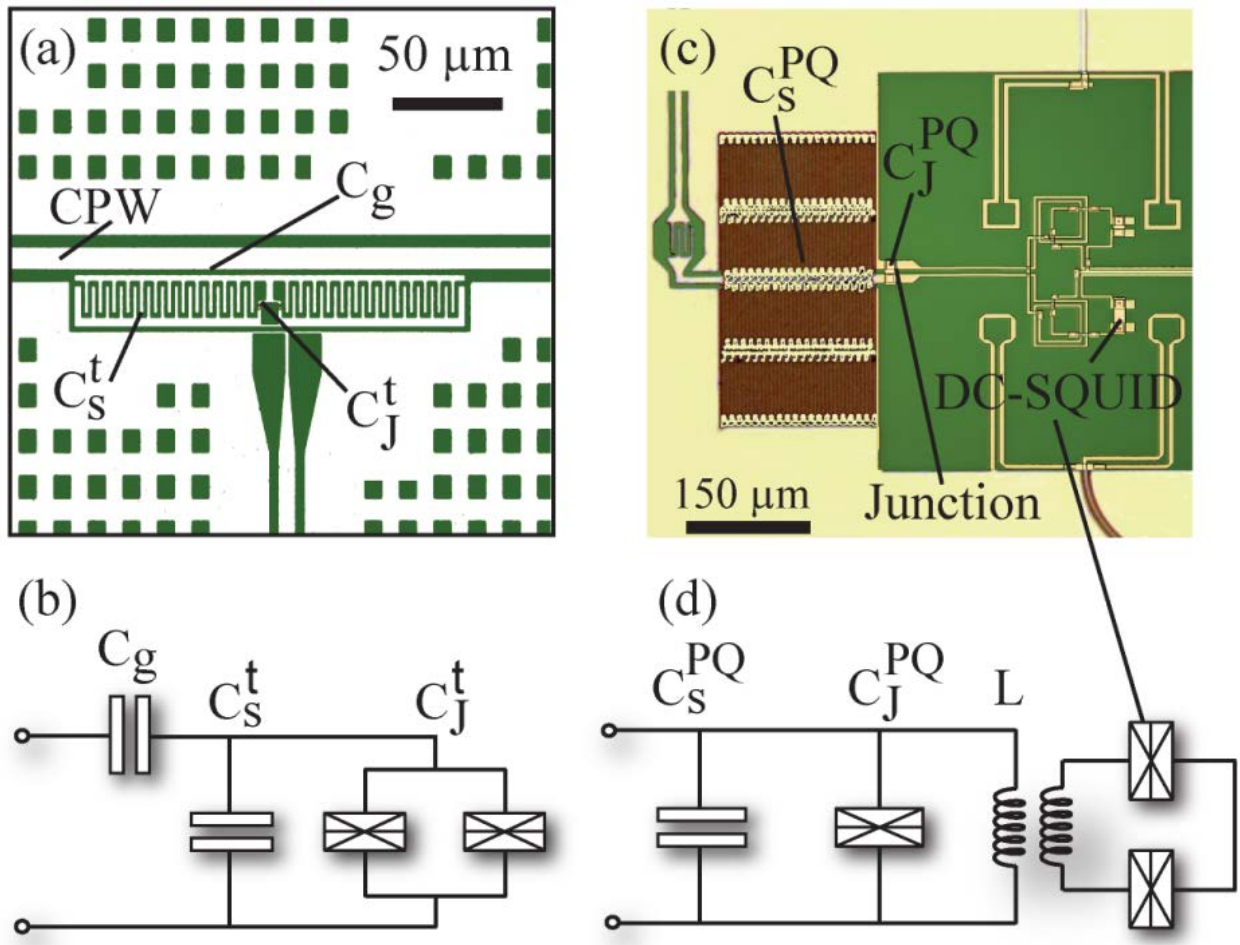

**Figure 2.** In (a) the optical microscope image of a transmon qubit, showing both the CPW resonator and the coupled JJ qubit. The schematic for the transmon qubit is drawn in (b). In (c) and (d) the same illustrations is presented for the phase qubit layout and its schematics.

Five different devices - two CPWs, two PQs and one transmon - with different parameters have been measured to get information about the losses. The measurements for the devices have been done at different frequencies in the range 3 to 10 GHz at the base temperature (~30 mK) of a dilution refrigerator. All measurements have been performed with low energy excitations with $V_{rms} < 1\,\mu V$.

The CPW resonators were designed as 1.26 cm long transmission lines of 50 Ω characteristic impedance, terminated with $C_c$=5 fF coupling capacitors at both ends. They were characterized by transmission measurements in order to extract resonance peak linewidth that defines the loaded $Q$ value, $Q_L$. The CPW resonators are under coupled and the intrinsic $Q$ value, $Q_0$, is determined by $Q_0 = Q_L/(1 + S_{21}(f_c))$ where $S_{21}(f_c)$ is the measured insertion loss [19].

For the qubits we have measured the excited state lifetime through the decay time after π-pulse Rabi excitations [6]. The transmon measurement was realized by a 2-tone excitation technique, where both a measurement signal and the Rabi pulse were sent through the CPW resonator. The read-out of the transmon was done in the weak dispersive limit [20], with CPW coupling $g/2\pi \sim 40$ MHz and CPW detuning $\Delta \sim 2$ GHz. For the phase qubit measurements, the Rabi excitation was applied with a flux bias line which is inductively coupled to the JJ loop (figure 2(c) and (d)). The read-out was done by applying a short read-out flux pulse to the JJ, which discriminates between the ground and excited state of the qubit. This long lived state was then measured with the inductively coupled dc SQUID. The phase qubit energy level spectroscopy data in figure 3, was obtained by sweeping the frequency of the applied flux bias. The spectroscopy data illustrates the single-photon transitions between the qubit ground and excited state. A more thorough analysis of hybrid TLS-qubit systems, from spectroscopy and time-domain measurements, can be found in Refs. [21,22].

**Table 1.** Sample parameters. The parameters that are not applicable are marked with 'X', and non-relevant ones with '–'. The tunnel junction areas $A_j$ indicate the nominal values and the tunnel junction resistance $R_j$ is measured from an on-chip test junction. The values of the Josephson coupling energy, $E_J$, are extracted from $R_j$ through the Ambegaokar-Baratoff relation [23]. Total capacitances, $C_\Sigma$, are estimated from the circuit designs. $C_\Sigma$ can also be estimated from frequency spectroscopy measurements, with the aid of estimates of $E_J$. The error bounds for $C_\Sigma$ when using $E_J$ for the calculation is hard to estimate compared to the geometrical capacitances, which we have used here.

|  | $A_j$ (μm$^2$) | $R_j$ (kΩ) | $C_\Sigma$ (fF) | $E_C/h$ (GHz) | $E_J/h$ (GHz) | $f_c$ (GHz) | $f_{01}$ (GHz) | $T_1$ (ns) | $Q$ | $\tan\delta$ |
|---|---|---|---|---|---|---|---|---|---|---|
| CPW A | x | x | - | x | x | 5.09 | x | x | 15593 | 6.2e-5 |
| CPW B | x | x | - | x | x | 5.02 | x | x | 16225 | 6.4e-5 |
| Transmon A | 1.0 | 4.2 | 97 | 0.21 | 33 | 5.0 | 3.01 | 131 | 2474 | 4.0e-4 |
| PQ-A | 1.7 | 0.14 | 697 | 0.027 | 1080 | x | 7.86 | 31 | 1531 | 6.5e-4 |
| PQ-B | 1.7 | 0.16 | 581 | 0.033 | 870 | x | 10.3 | 130 | 8413 | 1.2e-4 |

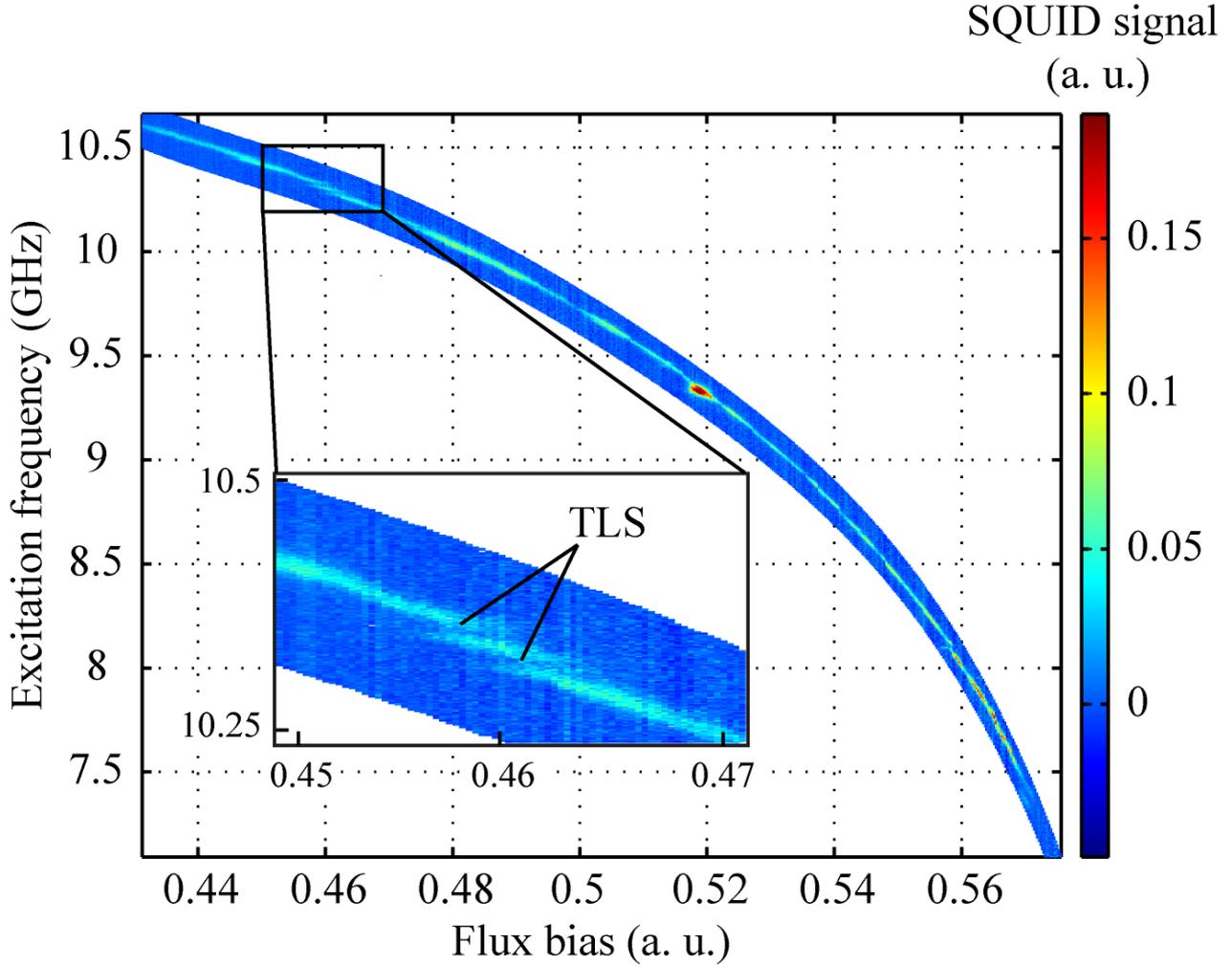

**Figure 3.** Spectroscopy of phase qubit PQ-A. In the enlargement, two avoided level crossings due to single-photon transitions between the qubit-TLS are indicated.

### 3. Results and discussion

*3.1. Relaxation times and quality factors*

*CPW resonators:* For the two measured CPW resonators we get an $Q_0 \sim 16\,000$, at resonance frequency $f_c \sim 5$ GHz (see table 1). The losses agree well with what has been obtained in earlier studies [13] for double layered substrates with Si and 300 nm thermal $SiO_2$, where the resonators are assumed to be limited by the dielectric losses of the substrate.

*Transmon:* The excited state population as a function of delay time, after a π-pulse from the ground state, is shown in figure 4. The relaxation time, $T_1$, was measured to be $T_1$=131 ns at an energy level separation corresponding to $f_{01} = 3$ GHz . This is well below the expected $T_1$ for this design of qubit [4, 5], with regard to CPW resonator-qubit coupling and influence from control and back-action, which limits the $T_1$ to a few μs. We attribute the limiting factors to dielectric material losses, also in agreement with other multi-layer processing of transmon qubits [24]. Indeed, typical transmon qubits with long $T_1$ are fabricated on low loss sapphire substrate, with ultra-small tunnel junctions and with no deposited dielectric layers [5].

*Phase qubits:* The decay time measurements for the two phase qubit samples, PQ-A and PQ-B, gave $T_1$ =31 ns at $f_{01} = 7.9$ GHz and $T_1$ =130 ns at $f_{01} = 10.3$ GHz, respectively (see figure 4). Here the large improvement for PQ-B is attributed to the removal of the deposited $SiN_x$.

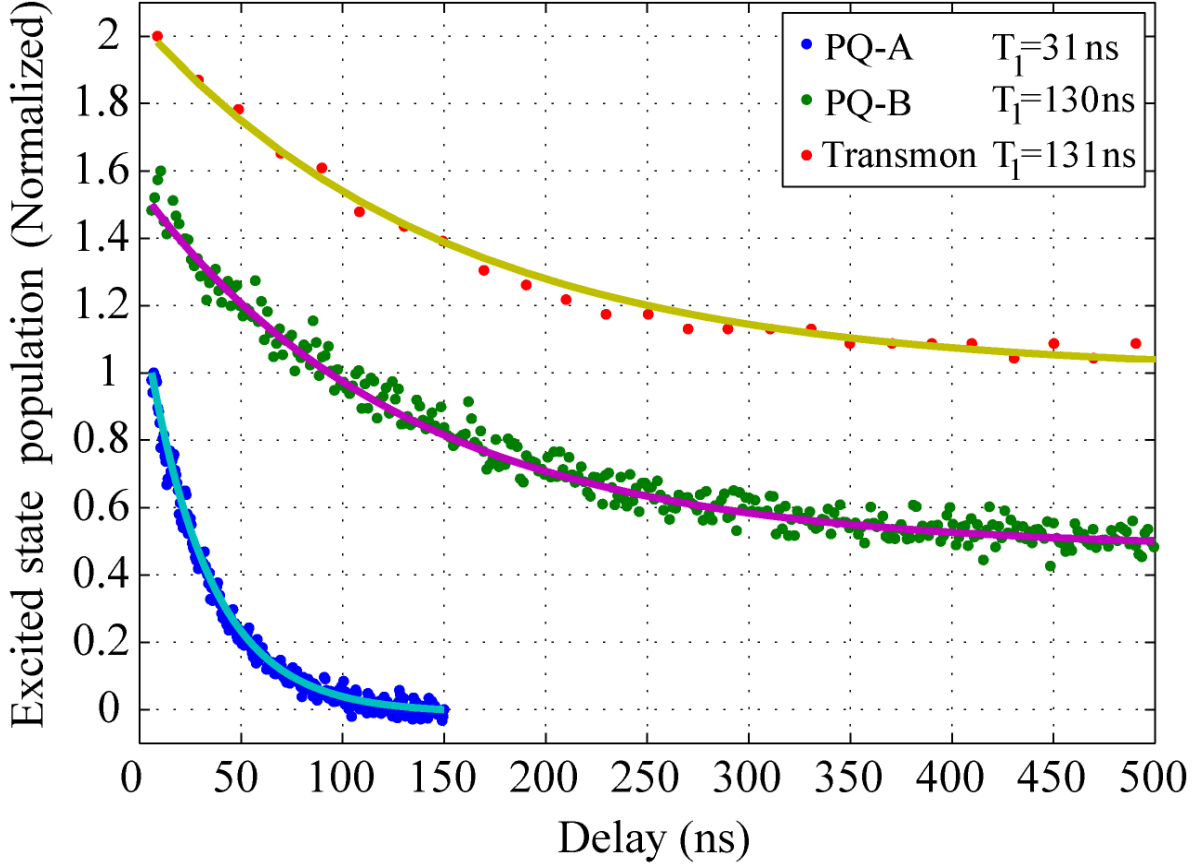

**Figure 4.** Excited state population as a function of delay time after π-pulse for the three qubit samples. Solid lines denote the corresponding fits to exponential decay. The curves are shifted in y-direction for clarity.

*3.2. Loss estimation*

We use the model for the dielectric losses, based on electrical dipole TLS residing in the dielectrics to evaluate our qubits [12, 17]. An example of this loss mechanism can be found in the enlargement of the spectroscopy data in figure 3, as the avoided level crossings from qubit-TLS interaction. It is valuable to get an upper estimate on how much the different dielectrics, $AlO_x$, $SiN_x$, and $SiO_2/Si$, contribute to the total loss of the qubits. Since the coupling is through electric field, the main loss is due to the electric field inside a dielectric or, equivalently, electric energy stored on the capacitive elements of the devices. The dielectric losses for a capacitor can be expressed with a complex permittivity $\varepsilon = \varepsilon_1 - i\varepsilon_2$, where $\tan \delta = \varepsilon_2/\varepsilon_1$. The qubits can be modelled as parallel LC-resonators, where the nonlinear inductance and capacitance are defined through the Josephson junction and the geometrical surroundings. At resonance the electric energy of the resonator is $E_e = \frac{|V_{rms}|^2}{4} C_\Sigma$. In table 1, the relaxation times for our devices have been recalculated into equivalent tangent loss factors, through the relations $Q = 2\pi f_{01} T_1$ and $\tan \delta = 1/Q$ [12]. By weighting the contributions from the different dielectrics to the total loss, with their respective capacitance fraction of the total capacitance, we can extract the materials loss factors [24]. The total loss can be written as

$$\tan \delta = \sum_i P_i \tan \delta_i ,$$

where $P_i = C_i/C_\Sigma$, is the capacitance fraction for the i:th included dielectric, $C_i$ and $\tan \delta_i$ its respective capacitance and loss tangent, and $C_\Sigma$ the total capacitance.

Our qubits contain both IDC and plate capacitors (tunnel junctions) geometries (see figure 2 (c) and (d)). To get the individual capacitances we have estimated them from well-known geometrical methods. For the IDC's we have used the method of conformal mapping to get the capacitance on multi-layer dielectric stacks [25, 26]. Especially for the IDCs, we have used this method to estimate how large fraction of the electric energy is stored in the covering $SiN_x$ layer and the substrate, respectively. The dielectric thicknesses are $t_{SiOx} = 300$ nm and $t_{SiNx} = 250$ nm and we have used dielectric constants $\varepsilon_{SiN_x} \simeq 7.5$, $\varepsilon_{SiO_2} \simeq 3.9$ and $\varepsilon_{Si} \simeq 11.9$. 19 % of the energy of the IDC is stored in the $SiN_x$, 71 % in the substrate $SiO_x$ and Si stack and the reminder in vacuum. $SiN_x$ also plays a role in the transmon, and from a similar estimation the energy stored in the IDC's $SiN_x$ is ~7% . The capacitances for the tunnel junctions we have estimated with a value of 60 fF/µm$^2$, which is the value obtained from measurements on 0.5-2 µm$^2$ tunnel junctions made with the same process [18]. We have excluded capacitances that contribute < 0.1 % to $C_\Sigma$. The included capacitances and the material weights are presented in table 2. The CPW resonators are made in the last steps of processing, and we assume them to be limited by the losses from the substrate dielectrics.

**Table 2.** Capacitances and weight factors. The qubits consist of two main capacitors, the tunnel junction capacitance, $C_j$, and a shunt capacitance, $C_s$. The top part of the table shows the estimated capacitances for each device. The lower part shows the weights for the different dielectrics. The weight for $SiN_x$ comes from the fraction of energy stored in the $SiN_x$ surrounding the IDC shunts, $C_s$. For PQ B, the $SiN_x$ was removed by post-processing (see Sect. 2)

| Capacitance | Transmon | PQ A | PQ B | CPW A & B |
|---|---|---|---|---|
| Tunnel Junction | $C_j$=60 fF | $C_j$=101 fF | $C_j$=101 fF | x |
| Shunt IDC | $C_s$=26 fF | $C_s$=596 fF | $C_s$=480 fF | x |
| Coupling | $C_g$=11 fF | | | x |
| $P_{AlOx}$ | 61 % | 14 % | 17 % | 0% |
| $P_{SiNx}$ | 3 % | 20 % | 0% | 0% |
| $P_{sub}$ | 36 % | 66 % | 83 % | 100% |

The fitting procedure is a least square method of the linear system of equations that has the form of $\mathbf{A}\,\mathbf{x} = \mathbf{b}$, with $\mathbf{A}$ being the matrix of material weights, $\mathbf{b}$ is a vector with elements of the measured loss and $\mathbf{x}$ is the fitted vector of tan δ for the dielectrics of our process:

$$\overbrace{\begin{bmatrix} P_{AlOx}^{CPWA} & P_{Sub}^{CPWA} & P_{SiNx}^{CPWA} \\ P_{AlOx}^{CPWB} & P_{Sub}^{CPWB} & P_{SiNx}^{CPWB} \\ P_{AlOx}^{PQA} & P_{Sub}^{PQA} & P_{SiNx}^{PQA} \\ P_{AlOx}^{PQB} & P_{Sub}^{PQB} & P_{SiNx}^{PQB} \\ P_{AlOx}^{t} & P_{Sub}^{t} & P_{SiNx}^{t} \end{bmatrix}}^{\mathbf{A}} \overbrace{\begin{bmatrix} \tan\delta_{AlOx} \\ \tan\delta_{Sub} \\ \tan\delta_{SiNx} \end{bmatrix}}^{\mathbf{x}} = \overbrace{\begin{bmatrix} \tan\delta_{CPWA} \\ \tan\delta_{CPWB} \\ \tan\delta_{PQA} \\ \tan\delta_{PQB} \\ \tan\delta_{t} \end{bmatrix}}^{\mathbf{b}}. \quad (1)$$

We include all data from the qubits and the CPW resonators in the fit. In table 3, the solution $\mathbf{x}$, tan $\delta$ for the dielectrics, are presented and we find that the losses are in good agreement with results from similar measurements from other groups [13, 14, 24].

**Table 3.** Tangent loss parameters of different materials obtained from the fitting. The errors arise from the geometrical uncertainties in our capacitance estimates, and represents the worst case from propagating errors.

| Material | tan $\delta$ |
|---|---|
| AlO$_x$ | 4.8e-4 ± 2e-4 |
| SiN$_x$ | 30e-4 ± 5e-4 |
| Substrate | 0.6e-4 ± 0.3e-4 |

This analysis is based on the estimated values of the capacitances with errors that can be quantified. The uncertainty in the lateral dimensions of our process is < 20 %, which will introduce errors in the capacitance estimates for both the IDC's and tunnel junctions. For the tunnel junctions, additional uncertainty comes from using the measured junction capacitances from similar devices, which we estimate to be <40%. We have made an error analysis by allowing an input uncertainty of the capacitance values according to the dimension uncertainties from the process. In table 3, the resulting propagating errors to the fit are included. The uncertainties fall within reasonable limits and do not change the conclusions from our analysis. A comparison of how the different dielectrics contribute to the total loss can be found table 4. The sum of each device's loss contributions in table 4, agrees well with the measured losses (see table 1). It is important to point out that the results from the different samples have been measured at different frequencies (see table 1), but the distribution in the number of TLS and coupling strengths to the qubits has been reported to be fairly uniform in this frequency range [12, 21].

The number and strength of avoided level crossings seen in the spectroscopic data, see figure 3, can be used to compare our results from relaxation time measurements [12]. We find a defect density $\sigma h \sim 2.4$ (GHz$\mu$m$^2$)$^{-1}$ and the maximum level splitting $S_{max}/h \sim 0.02$ GHz from samples PQ-A and PQ-B and this gives a lifetime, $T_1 \sim 185$ ns. Here we have used the relation $1/T_1 = (\pi/6)\sigma A S_{max}^2/\hbar$ derived in [12], which describes the lifetime in tunnel junctions from the strength and density of avoided level crossings. Translated into dielectric loss factors for the tunnel junctions in PQ-A and PQ-B we get tan $\delta_{AlOx}$ =1.1e-4 and tan $\delta_{AlOx}$ = 8.4e-5, respectively. Since the phase qubits seems to be limited by the additional losses to the SiN$_x$ and substrate according to our analysis, we compare these values to the weighted loss tangents for AlO$_x$ found in table 4. For PQ-B it is in very good agreement with the spectroscopic data, while there is a larger discrepancy for PQ-A. This is explained by the dominant loss to SiN$_x$ for PQ-A compared to the almost equal loss to the substrate for PQ-B, which results in a greater error for the loss to AlO$_x$ in PQ-A.

**Table 4.** The fitted loss attributed to the different dielectrics. The limiting losses for the different samples are indicated in bold letters. The losses for the two CPW resonators are the same, since all loss is assumed to be in the substrate.

| | $P_{AlOx}$ tan $\delta_{AlOx}$ | $P_{SiNx}$ tan $\delta_{SiNx}$ | $P_{sub}$ tan $\delta_{sub}$ | $\Sigma P_x$ tan $\delta_x$ |
|---|---|---|---|---|
| PQ-A | 6.9e-5 | **5.5e-4** | 3.4e-5 | 6.5e-4 ± 1.3e-4 |
| PQ-B | **8.4e-5** | 0 | **4.7e-5** | 1.3e-4 ± 4e-5 |
| Transmon | **3.0e-4** | 8e-5 | 2e-5 | 4.0e-4 ± 1.4e-4 |
| CPW A & B | 0 | 0 | **5.7e-5** | 5.7e-5 ± 3e-5 |

## 4. Summary and conclusions

By comparing the excited lifetimes and analysing the material losses, we were able to extract the contribution from different materials to the total loss of each device (see table 4). The result from the spectroscopic data also agrees well with our analysis, hence strengthens its validity. For PQ-A it is evident that the deposited $SiN_x$ on top of the IDC contribute to its lower performance compared to sample PQ-B where it has been removed. The lifetime of PQ-B is increased by more than a factor four, and is limited by the roughly equal losses coming from the $AlO_x$ and the shunting IDC on the substrate. This indicates that post-processing steps, which potentially could be harmful to the tunnel junction quality, actually improves the qubit lifetime. The best phase qubits to date, based on multilayer tunnel junction processing, have had the shunting IDC capacitor replaced by a plate capacitor with a low-loss dielectric made of a-Si:H. This gives roughly three times better lifetimes than reported here [27]. A similar analysis as we have done here, indicates that these qubits are not limited by the dielectric losses, and it is speculated that losses instead are due to non-equilibrium quasiparticles [28]. From the data in table 4, it can be seen that the transmon qubit is clearly limited by the tunnel junction $AlO_x$ quality, in its current design. One order of magnitude lower losses of micron-sized $AlO_x$ tunnel junctions were reported, with epitaxial grown $Al_2O_3$, where transmon relaxation time was measured to be $T_1=0.76$ µs [24].

We have provided support to the assumption that material losses from dielectric layers limit the coherence of multi-layer JJ qubits. At the same time, these qubit coherence measurements can also be understood as a way to characterize the losses in materials at high-frequency and low probe power. We have shown that this type of loss characterization works well to quantify the quality of our highly scalable process, even with data from different qubit types. Previous work on various superconducting qubits has also indicated that one source of decoherence are the losses in the dielectric materials used for fabrication. However, different qubits were fabricated with different technologies and by different laboratories, and this diversity has precluded so far a clear comparison. In this paper we present for the first time a comparison between two different types of qubits fabricated with the same technology, thus allowing us to present a consistent picture on the effects of material losses in multi-layer JJ quantum bits. The information from our analysis is helpful for improving future sample designs, where the limits of these materials can be approached. It should be possible for the phase qubits fabricated by this technology to reach similar values as in Refs. [27], where the qubits performance is limited by other loss mechanisms [28]. The number of TLS's is proportional to the junction dimensions, and by reducing the tunnel junction size, the losses due to $AlO_x$ can be reduced [12]. In combination with an increased IDC capacitance on optimized substrates, such as sapphire or low-loss Si [13], the phase qubit dielectric loss could be reduced by almost one order of magnitude, hence not being the limiting factor. Furthermore, for the transmon qubits, improvements of a factor of three could be obtained by using a low-loss substrate and reducing the tunnel junctions area to the limit of standard i-line projection lithography ∼0.5 µm. To approach similar quality as transmons fabricated with e-beam lithography with *in-situ* double angle evaporation, resulting $T_1 > 2$ µs, an improved process is needed. Such process should have improved $AlO_x$ quality, obtained for example by epitaxial $AlO_x$ [15, 24], or utilize e-beam lithography for smaller tunnel junction definitions.

**Acknowledgements**
This work has been financially supported by Academy of Finland (projects No 141559 and 135135), NGSPM, Tekes (project FinCryo No 220/31/2010), and European Research Council. J.-M.P. was financially supported by the Väisälä foundation.